\documentclass[preprint,showpacs,superscriptaddress,prd,nofootinbib]{revtex4}
\pdfoutput=1
\usepackage{mathrsfs}
\usepackage{amsfonts,amsmath,amssymb}
\usepackage{enumerate}
\usepackage{hyperref}
\usepackage{graphicx}

\newcommand{\bea}{\begin{eqnarray}}
\newcommand{\eea}{\end{eqnarray}}
\newcommand{\be}{\begin{equation}}
\newcommand{\ee}{\end{equation}}

\newcommand{\nn}{\nonumber}

\newcommand{\Tr}{\text{Tr}}

\newcommand{\cp}{\mathbb{CP}}

\newcommand{\ads}{{\rm AdS}}
\newcommand{\cft}{{\rm CFT}}
\newcommand{\cN}{{\cal N}}

\newcommand{\la}{\lambda}
\newcommand{\tla}{\tilde{\lambda}}
\newcommand{\Da}{\Delta_a}
\newcommand{\Db}{\Delta_b}
\newcommand{\Dq}{\Delta_f}
\newcommand{\Dqt}{\Delta_{\tilde f}}

\begin{document}
\title{Calculating the partition function of $\bf \cN=2$ Gauge theories on $\bf S^3$ and AdS/CFT correspondence}
\author{Sangmo Cheon}
\affiliation{Department of Physics and Astronomy \& Center for Theoretical Physics, \\
Seoul National University,   Seoul 151-747, Korea}
\author{Hyojoong Kim}
\affiliation{Department of Physics
and Research Institute of Basic Science, \\ Kyung Hee University,
 Seoul 130-701, Korea}
\author{Nakwoo Kim}
\affiliation{Department of Physics
and Research Institute of Basic Science, \\ Kyung Hee University,
 Seoul 130-701, Korea}
\affiliation{Department of Physics and Astronomy,
\\ University of British Columbia, Vancouver,
  British Columbia V6T 1Z1, Canada}


\begin{abstract}
We test the AdS/CFT correspondence by computing the partition function of some $\cN=2$ quiver Chern-Simons-matter
theories on three-sphere. The M-theory backgrounds are of the Freund-Rubin type with the seven-dimensional internal space
given as Sasaki-Einstein manifolds $Q^{1,1,1}$ or $V^{5,2}$. Localization technique reduces the exact path integral
to a matrix model, and we study the large-$N$ behavior of the partition function.
For simplicity we consider only non-chiral models which have a real-valued partition function.
 The result is in full agreement with
the prediction of the gravity duals, i.e. the free energy is proportional to $N^{3/2}$ and the coefficient matches
correctly the volume of $Q^{1,1,1}$ and $V^{5,2}$.
\end{abstract}

\pacs{11.15-q,11.25.-w}

\maketitle


\section{Introduction}
We have seen remarkable progress made about M2-brane dynamics recently. It is now well established that $\cN=6$ Chern-Simons gauge field theory (ABJM model) with $U(N)_k\times U(N)_{-k}$ correctly describes M2-branes at orbifold singularity $\mathbb C^4/\mathbb Z_k$ \cite{Aharony:2008ug}. But as usual with AdS/CFT correspondence, many of the checks with maximally possible duality relation strongly depend on the symmetry, and for a more convincing and challenging test, one would like to study models with less supersymmetries.

Indeed, there have been a number of works which put forward new, more nontrivial duality relations regarding M2-branes \cite{Martelli:2008rt,Martelli:2008si,Hanany:2008cd,Ueda:2008hx,Imamura:2008qs,
Hanany:2008fj,Franco:2008um,Aganagic:2009zk,Martelli:2009ga,Jafferis:2009th,Benini:2009qs}. 
The richest and still challenging category is probably $\cN=2$ models. The dual geometry is usually of Freund-Rubin type, i.e. $\ads_4\times X^7$ where $X^7$ is a seven-dimensional manifold satisfying the Sasaki-Einstein property. One can use the machinary of toric geometry \cite{Martelli:2004wu}, brane tiling \cite{Franco:2005sm}, crystals \cite{Lee:2006hw} etc., but we are yet to have a complete understanding even with the more tractable class of toric Sasaki-Einstein manifolds. Most of the earlier works just checked that (a particular branch of) the vacuum moduli space coincides with the gravity side, and studied the chiral ring structure perhaps including some nonperturbative states involving baryonic operators. 

The situation has changed recently, mainly due to important progresses on the field theory side. One is the computation of supersymmetric indices, where one can certainly check beyond the chiral ring. Indeed, some of the models which showed precise vacuum moduli space turned out to give wrong indices and are ruled out. For more details, see \cite{Bhattacharya:2008bja,Kim:2009wb,Imamura:2009hc,
Imamura:2010sa,Kim:2010vwa,Imamura:2011uj,Cheon:2011th}. The other route is the computation of exact partition function on three-sphere, and it is the aim of this paper to present calculation of the partition function for some $\ads_4/\cft_3$ proposals with four supercharges.

It has been known for a long time that in M2-brane dynamics the number of degrees of freedom scale as $N^{3/2}$, where $N$ is the number of M2-branes. First of all, using localization technique it is shown that the partition function on three-sphere can be expressed as a matrix integral \cite{Kapustin:2009kz}.  Using an exact resolvent 
it 
was first confirmed in a beautiful paper \cite{Drukker:2010nc} that the free energy has the right behavior of $N^{3/2}$. For more details and related works, readers are referred to 
\cite{Pestun:2007rz,Marino:2009jd,Halmagyi:2003ze,Suyama:2010hr}.

For more general models, we  do not have the luxury of an exact resolvent, but a powerful numerical and analytic method have been proposed in \cite{Herzog:2010hf}. In that procedure, the matrix integral is approximated using saddle point method. The saddle point equations are turned into an auxiliary first-order coupled differential equations which quickly give the information on the behavior of roots. In \cite{Herzog:2010hf} the authors were interested in tri-Sasakian geometry and $\cN\ge 3$ quiver Chern-Simons models. For the models studied in \cite{Herzog:2010hf}, the free energy exhibits nice $N^{3/2}$ behavior  and the coefficient is consistent with the volume of seven-dimensional internal space through the well-known AdS/CFT dictionary
\be
F = N^{3/2} \sqrt{\frac{2\pi^6}{27 \rm{Vol}(X^7)}} \,\,. 
\label{dict}
\ee

In these models with $\cN\ge3$, all the fields are equipped with canonical conformal dimension and R-charge, i.e. $\Delta=R=1/2$. But in $\cN=2$ theories in general the fields can acquire anomalous dimensions. The problem of putting general superconformal $\cN=2$  theories on three-sphere and computing the partition function has been solved in \cite{Jafferis:2010un,Hama:2010av}. More concretely, it is shown that chiral multiplets with an arbitrary R-charge can couple to gauge fields and curvature of $S^3$ in a way preserving $OSp(2|2)\times SU(2)$. This symmetry is a {\it half} of the superconformal group in addition to the isometry of $S^3$, $SU(2)\times SU(2)$.

Quoting eq.(1.2) of \cite{Jafferis:2010un}, the partition function is given as 
\be
Z = \int \prod_{Cartan} du \,\,  e^{i\pi \Tr \, u^2} \mbox{det}_{Ad} (\sinh(\pi u)) 
\prod_{R_i} {\rm{det}}_{R_i} ( e^{\ell(1-\Delta_i+iu)} ) , 
\ee
where $\Delta_i$ is the conformal dimension of $i$-th chiral multiplet in gauge group representation $R_i$. The function $\ell(z)$ is known as double sine function, see for more details  \cite{Kharchev:2001rs,Bytsko:2006ut}.

One can in principle try to compute $Z$ for all the Chern-Simons dual proposals. When we started working with them, we have found that chiral-models which give complex-valued $Z$ are rather difficult to deal with. We do not have the full understanding of how the roots condense in large-$N$ limit yet. On the other hand, it turns out that simpler non-chiral models with two-nodes produce dependable but still nontrivial results. In other words, we can verify the free energy agrees with \eqref{dict} in large-$N$ limit.

In this paper we study three $\ads_4/\cft_3$ examples.  In Sec.\ref{adj} we study the first example which is dual to $V^{5,2}$, where the gauge theory comes with adjoint representation. In Sec.\ref{fundam} we study models involving chiral multiplets in fundamental representations. One is dual to $Q^{1,1,1}$, and the last example provides an alternative description of $V^{5,2}$. 
We give a brief discussion on our results in Sec.\ref{discussion}. 

{\bf Note added:} There are two papers on the arxiv which have significant overlap with this article. As this paper was being typed, we became aware of a work \cite{Martelli:2011qj} and we  coordinated the release of the paper. Later we received \cite{Jafferis:2011zi}, which provides a comprehensive analysis for non-chiral ${\cN=2}$ models. In particular, the F-theorem in \cite{Jafferis:2011zi} clarifies a misconception on $Z$-extremization in the original version of this paper. 
\section{Quiver with adjoint matters: $V^{5,2}$}
\label{adj}
In this section we consider the dual field theory for $V^{5,2}$ geometry. This geometry is a direct higher-dimensional generalization of conifold. The cone over $V^{5,2}$ is a singular space defined as
\be
z^2_1 + z^2_2 + z^2_3 + z^4_4+ z^2_5 = 0 , \quad z_i \in \mathbb C .
\ee
Obviously this geometry has $U(1)_R\times SO(5)$ isometry. $V^{5,2}$ is an example of non-toric Sasaki-Einstein manifolds. According to AdS/CFT, M-theory in $\ads_4\times V^{5,2}$ is dual to a strongly coupled three-dimensional $\cN=2$ conformal field theory.

\begin{figure}
\centering
\includegraphics[scale=0.35,trim= 20 20 40 40,clip=true]{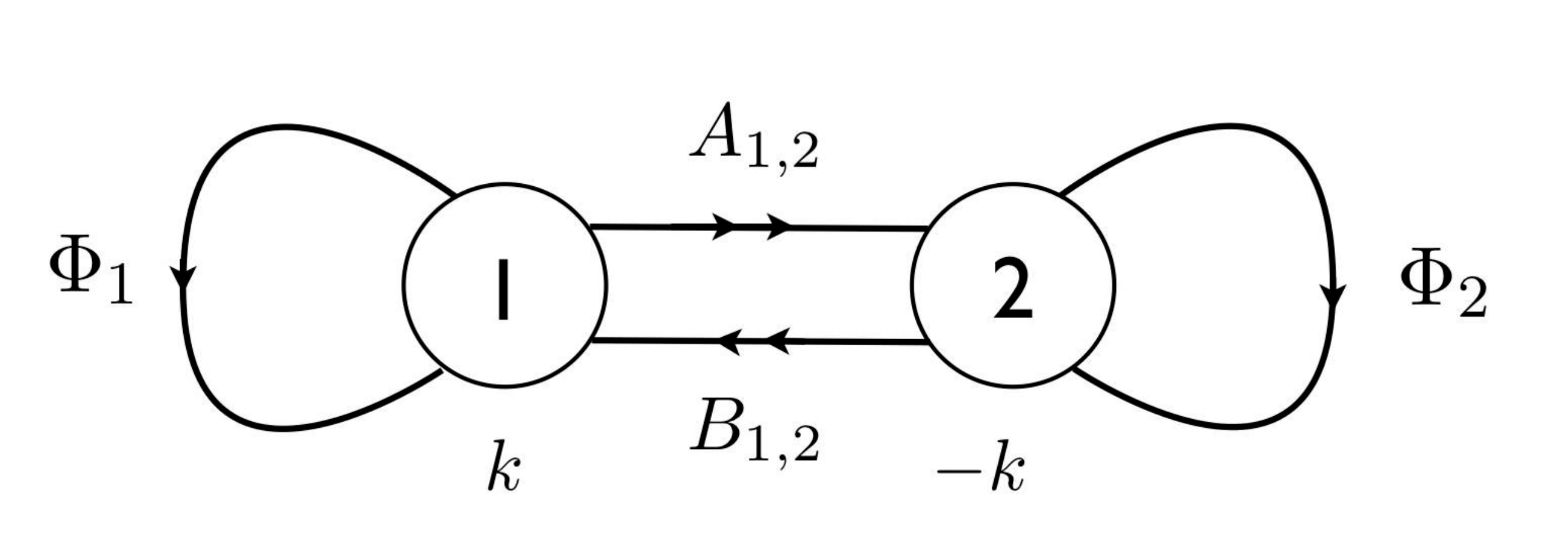}
\caption{The quiver diagram for Chern-Simons dual of $AdS_4 \times V^{5,2}$}
\label{quiver_v52a}
\end{figure}

The dual as a quiver Chern-Simons system is proposed in \cite{Martelli:2009ga}, see Fig.\ref{quiver_v52a}. The gauge group is $U(N)\times U(N)$ with Chern-Simons levels $(k,-k)$. The matter fields include six chiral multiplets in total: $A_a, \,a=1,2$ in $(N,\bar N)$ representation, $B_a,\, a=1,2$ in $(\bar N,N)$, and we also have one chiral multiplet in the adjoint representation for each gauge group, called  $\Phi_1,\Phi_2$. The vacuum moduli space provides an abelian orbifold $V^{5,2}$ when one chooses the following cubic superpotential
\be
W = \Tr (\Phi^3_1 + \Phi^3_2 + \epsilon^{ij} (\Phi_1 A_i B_j + \Phi_2 B_i A_j )) .
\label{sp1}
\ee
For a study of this duality relation using rotating membrane solutions, see Ref.\cite{Lee:2010hjb}.

It is straightforward to write down the formula for the partition function using the recipe given in \cite{Jafferis:2010un, Hama:2010av}, with arbitrary R-charges or equivalently the conformal dimensions of the matter fields.  From the symmetry of our system, it is obvious that the bifundamental fields have the same R-charges, and we will call it $\Db$. The two adjoint fields $\Phi_1,\Phi_2$ should also come with the same R-charge $\Da$. Now if we designate a Cartan subgroup for $U(N)\times U(N)$ using eigenvalues $\lambda_i,\tilde\lambda_i,\, i=1,\cdots,N$,  the partition function is given by the following integral.
\bea
Z &=& \frac{1}{(N!)^2}
\int \prod_{i=1}^{N} \frac{d\lambda_i d\tilde\lambda_i}{(2\pi)^2}
\exp{\left[\frac{ik}{4\pi}\sum_i (\lambda^2_i-\tilde{\lambda}^2_i)\right]}
\prod_{i<j}\left(2\sinh \frac{\la_i-\la_j}{2}\right)^2 \left(2\sinh \frac{\tla_i-\tla_j}{2}\right)^2
\nn\\
&&
\times \prod_{i,j} \exp{
\left[ \ell(1-\Da+i\frac{\la_i-\la_j}{2\pi}) +
\ell(1-\Da+i\frac{\tla_i-\tla_j}{2\pi}) \right] }
\nn\\
&&
\times \prod_{i,j} \exp{
\left[ 2\ell(1-\Db+i\frac{\la_i-\tla_j}{2\pi}) +
2\ell(1-\Db-i\frac{\la_i-\tla_j}{2\pi}) \right] }  .
\label{mi1}
\eea

The function $\ell(z)$ was used in \cite{Jafferis:2010un, Hama:2010av} to express a one-loop determinant for a chiral multiplet with a generic value of R-charge. As an infinite product, it is defined as
\be
\label{elldef}
\exp \ell(z) = \prod_{n=1}^{\infty} \left(\frac{n+z}{n-z}\right)^n .
\ee
More concretely it can be expressed as \cite{Jafferis:2010un}
\be
\ell (z) = -z \ln ( 1 - e^{2\pi i z}) + \frac{i}{2}
\left(
\pi z^2 + \frac{1}{\pi} \mbox{Li}_2(e^{2\pi i z}) \right)
-\frac{i\pi}{12} .
\label{ell}
\ee
Although it is not immediately seen from this expression, from the definition as infinite product it is clear $\ell$ is an odd function, and also $\ell(z)\in {\mathbb  R}$ if $z\in  {\mathbb  R}$.

For the manipulation of the integrand for $Z$, we find the following relation of $\ell(z)$ very useful. If $z=x+iy$ with $y\in \mathbb R$, we can derive
\bea
\lefteqn{\ell(x+iy) + \ell(x-iy) = \ell(x+iy)-\ell(-x+iy)}
\nn\\
&=& -2\pi x |y| +
\sum_{n=1}^{\infty}
e^{-2\pi n |y|}
\left[
\frac{2x \cos (2\pi n x)}{n}
-\left(\frac{2|y|}{n}+\frac{1}{\pi n^2}\right)\sin(2\pi n x)
\right] .
\label{elex}
\eea

Let us check the convergence of multiple indefinite integrals in \eqref{mi1}. First of all, the Chern-Simons part with $k$ dependence is highly oscillatory, but in itself the integral is convergent, since for instance $\int^\infty_{-\infty} \sin x^2 = \sqrt{\pi/2}$. Recall that we are doing the integration along real-axis. The remaining part in the integrand with $\sinh$ is exponentially growing, and the part with $\ell$ is exponentially decaying, as we send $\lambda$ or $\tla$ to infinity. Suppose we take one particular eigenvalue $\la_i$ to infinity, and keep others finite. Then the integrand, apart from the oscillating part, behaves like
\be
\exp \left[ (N-1)(\Da+2\Db-2)\la_i
\right]
\label{asym1}
\ee
Then it is obvious that the integration in \eqref{mi1} is divergent if $\Da+2\Db>2$. When  $\Da+2\Db<2$ it is absolutely convergent, and we expect $Z$ becomes smaller as we make R-charges smaller and smaller. We are interested in the marginal case of  $\Da+2\Db=2$ in this paper, but will make some comments about $\Da+2\Db<2$ case later. 

We are interested in the behavior of $Z$ in the leading order as $N\rightarrow\infty$. We will employ the saddle point approximation and take the large-$N$ limit, closely following the procedure described in \cite{Herzog:2010hf}. The saddle points are determined by setting the following quantities to zero.  In the following  $F$ is related to $Z$ by $Z=\int e^{-F}$.
\bea
\frac{\partial F}{\partial\lambda_i }&=&
-\frac{ik}{2\pi}\la_i - \sum_{j  \neq i}\coth\frac{\la_i-\la_j}{2} \nn \\
&+&\sum_j \Big[2m(1-\Db, \frac{\la_i-\tla_j}{2\pi})+m(1-\Da, \frac{\la_i-\la_j}{2\pi})\Big],
\label{spe1}
\\
\frac{\partial F}{\partial\tilde\lambda_i} &=&
\frac{ik}{2\pi}\tla_i - \sum_{j  \neq i}\coth\frac{\tla_i-\tla_j}{2} \nn \\
&-&\sum_j \Big[2m(1-\Db, \frac{\la_i-\tla_j}{2\pi})-m(1-\Da, \frac{\tla_i-\tla_j}{2\pi})\Big].
\label{spe2}
\eea
In order to avoid too much clutter we introduced a shorthand notation
\bea
m(x,y) &\equiv& -\frac{i}{2\pi}\left(\ell'(x+i y)-\ell'(x-i y)\right) \nn \\
&=& \frac{y \sin 2\pi x-x\sinh2\pi y}{\cos 2\pi x-\cosh 2\pi y} .
\eea
where we used
\be
\ell'(z) = -\pi z \cot(\pi z).
\ee

From the saddle point equations above, one easily notices that the saddle point equations are invariant under (i) $(\lambda_i,\tilde\lambda_i) \rightarrow
-(\lambda_i,\tilde\lambda_i)$, and (ii) $(\lambda_i,\tilde\lambda_i) \rightarrow
(\tilde\lambda_i^*,\lambda_i^*)$. The implication is that the root distribution is symmetric  with respect to the origin, and $\lambda_i$ is the complex conjugate of $\tilde\lambda_i$.

\begin{figure}
\centering
\includegraphics[scale=0.70,trim= 0 0 0 0,clip=true]{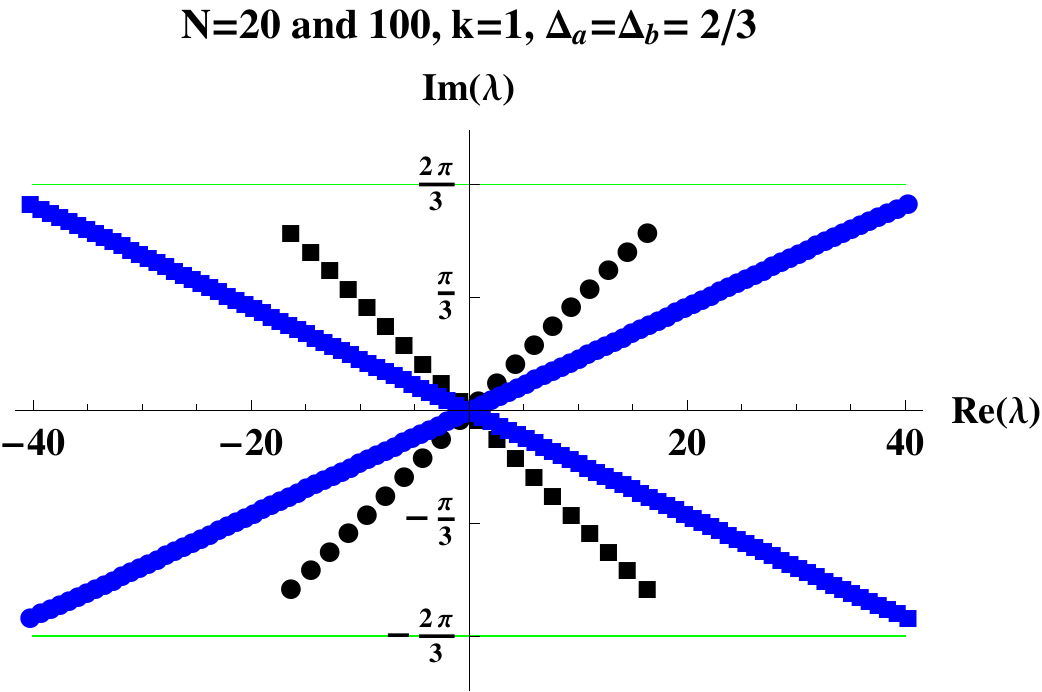}
\includegraphics[scale=0.70,trim= 0 0 0 0,clip=true]{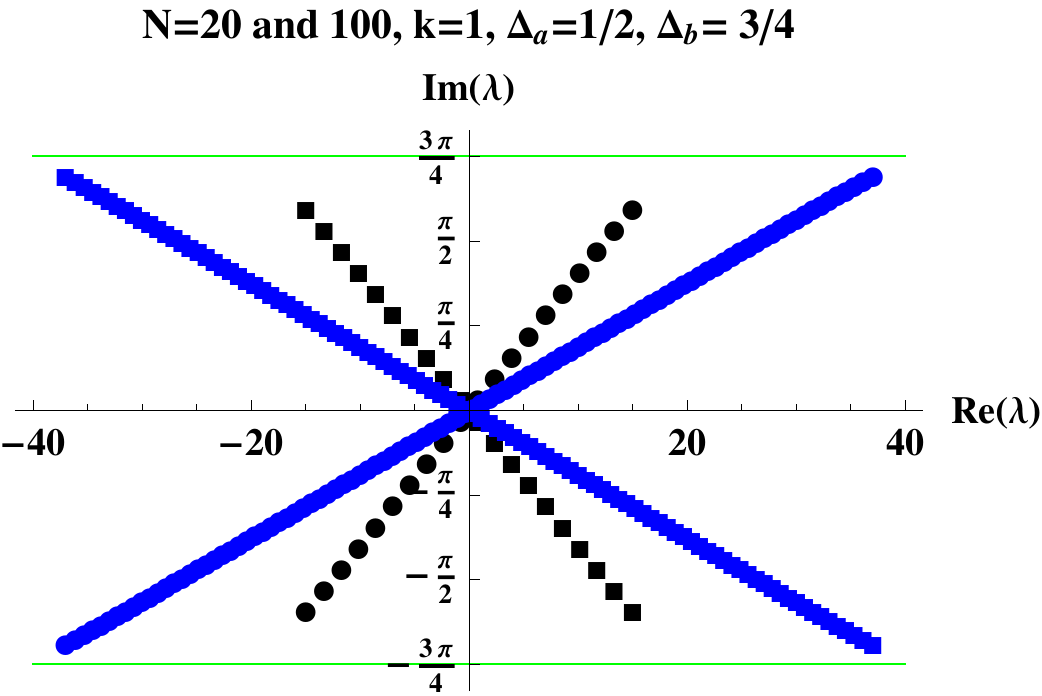}
\caption{These two figures show saddle points for $V^{5,2}$
theory with different R charges, from our numerical analysis.
The roots are plotted with black and blue dots for $N=20$ and $100$,
respectively. The real value of the roots scales as
$\sqrt{N}$. The  imaginary of part of the roots is 
bounded by $\pm\Delta_b \pi$, which are represented by green lines.}
\label{rootsV52}
\end{figure}

A practical method 
 to find the saddle points is to treat the eigenvalues $\lambda_i,\tilde\lambda_i$ as functions of {\it time} $t$, and solve an auxiliary  coupled ordinary differential equations (ODE)
\be
\frac{d\lambda_i}{dt} = \frac{\partial F}{\partial\lambda_i} \,  , \quad
\frac{d\tilde\lambda_i}{dt} = \frac{\partial F}{\partial\tilde\lambda_i} \,  .
\ee
By choosing the initial conditions $\lambda_i(0),{\tilde\lambda}_i(0)$ appropriately, the eigenvalues as functions of $t$ converge rapidly as $t\rightarrow\infty$. Obviously
the fixed points
$\lambda_i(\infty),{\tilde\lambda}_i(\infty)$ would give us the roots of saddle point equation. This method was utilized by the authors of Ref.\cite{Herzog:2010hf} to study the large-$N$ behavior of partition function for $\cN=3$ Chern-Simons quiver theories. We find this technique is dependable for the non-chiral models we deal with in this paper, and we closely follow the procedure in \cite{Herzog:2010hf}.

We put the coupled ODE on computer and have obtained the roots. They condense and make a cut, and the roots for some different values of R-charges are displayed on complex plane in Fig.\ref{rootsV52}. We first note that the distribution is symmetric under  $(\lambda_i,\tilde\lambda_i) \rightarrow
-(\lambda_i,\tilde\lambda_i)$, and in fact it always holds that
$\lambda_i^*=\tilde\lambda_i$.  This is of course not unexpected, considering the symmetry of the integral mentioned earlier. 

As we analyze the numerical data for different values of $N$, we find that  the real part of $\lambda$ scales as $N^\alpha$ with $\alpha \approx 1/2$, and the imaginary part stays within the same interval of order 1.  This behavior is qualitatively the same as the case of ABJM model reported in \cite{Herzog:2010hf}, and we also write
\be
\lambda_i = N^{\alpha} x_i + i y_i .
\label{laxy}
\ee
For convenience, we assume we have re-arranged the roots so that $x_i<x_j \mbox{ if } i<j$. As $N$ goes to infinity, $x_i, y_i$ become continuously distributed over an interval.

Another notable feature of Fig.\ref{rootsV52} is that the range of $\mbox{Im} (\la_i)$ changes for different values of R-charge. This can be understood if we go back to the saddle point equations \eqref{spe1},\eqref{spe2}. Let us look at the following term from the second line of \eqref{spe1}.
\be
\sum_j  m(1-\Db, \frac{\la_i-\tla_j}{2\pi}) . 
\ee
With \eqref{laxy}, for $j=i$ this expression contains 
\be 
m(1-\Db, i y_i/\pi) = 
\frac{i}{\pi} \frac{y\sin(2\pi(1-\Db))-\pi(1-\Db)\sin(2y)}{\cos(2\pi(1-\Db))-\cos(2y)} \, , 
\label{force}
\ee
which is divergent at $y=\pi \Db$. At this point the roots feel  {\it infinite force}, and the roots cannot cross this boundary. This  argument is verified in Fig.\ref{rootsV52}.

Now we take the large-$N$ limit of $F$ analytically. We first write $F=F_{ext}+F_{int}$, where $F_{ext}$ describes the Gaussian part with a single summation over the eigenvalues, and $F_{int}$ describes the {\it interaction} of the eigenvalues with a double summation.  Juggling with the summation indices, we can rewrite $F_{int}$ in the following way.
\bea
F_{int} &=& -\sum_{i<j} \Big\{
\ln e^{\la_j-\la_i} + \ln e^{\tla_j-\tla_i}
+ 2 \ln ( 1 - e^{\la_i-\la_j}) +2 \ln ( 1 - e^{\tla_i-\tla_j})
\nn\\
&&
+\ell(1-\Da+i\frac{\la_j-\la_i}{2\pi})-\ell(-1+\Da+i\frac{\la_j-\la_i}{2\pi})
\nn\\
&&
+\ell(1-\Da+i\frac{\tla_j-\tla_i}{2\pi})-\ell(-1+\Da+i\frac{\tla_j-\tla_i}{2\pi})
\nn\\
&&
+2\ell(1-\Db+i\frac{\tla_j-\la_i}{2\pi})-2\ell(-1+\Db+i\frac{\tla_j-\la_i}{2\pi})
\nn\\
&&
+2\ell(1-\Db+i\frac{\la_j-\tla_i}{2\pi})-2\ell(-1+\Db+i\frac{\la_j-\tla_i}{2\pi})
\Big\}
\nn\\
&&
+\sum_i \Big\{
2\ell(1-\Da)+2\ell(1-\Db+i\frac{\la_i-\tla_i}{2\pi})
+2\ell(1-\Db-i\frac{\la_i-\tla_i}{2\pi})
\Big\} .
\label{fint}
\eea

It turns out that the terms within the double summation give ${\cal O}(N^{2-\alpha})$ contribution, and the single summation part in the last line of \eqref{fint} is subleading as ${\cal O}(N)$. Now for the terms involving the function $\ell$ for lines 2-5 of \eqref{fint}, we can make use of the relation \eqref{elex} . One discovers that
 the linear terms such as $\ln e^{\lambda_j-\lambda_i}+ \ln e^{\tla_j-\tla_i} $ in the first line get exactly canceled by similar terms from the matter fields, if we impose
\be
\Da+ 2 \Db  = 2 .
\label{Dsum}
\ee
This is essentially the same asymptotic property we discussed around Eq.\eqref{asym1}.
It is certainly consistent with the superpotential \eqref{sp1}, since $\Tr (AB\Phi)$ should have R-charge 2.  In fact, considering everything in \eqref{sp1} one easily sees we should set $\Da=\Db=2/3$, if we are to relate this system to $V^{5,2}$. But we will just assume \eqref{Dsum} and leave $\Db$ undetermined in the following computation.

Now the rest of the summand is given as a series expansion, and we move on to
 take large-$N$ limit.
We introduce a continuous variable $s$ instead of summation indices $i,j$,
\be
s=i/N, \quad 0<s<1 .
\ee
Then a summation is transformed into an integral,
\be
\frac{1}{N}\sum_{i=1}^N F(i/N) \rightarrow  \int^1_0  F(s) ds
= \int^{x_*}_{-x_*} F(s(x)) \rho(x) dx ,
\ee
where $x_*=\mbox{Max}(x_i)$ and $\rho(x)=\tfrac{ds}{dx}$.

Ignoring the terms in the last line of \eqref{fint},  we obtain
\bea
\lefteqn{F_{int} = - N^2 \int^{x_*}_{-x_*} \rho(x) dx \int^x_{-x_*} \rho(x') dx'
\sum_{n=1}^{\infty} \left[
-\frac{2}{n} \left( e^{n(\la(x')-\la(x))} +e^{n(\tla(x')-\tla(x))} \right)  \right.
}
\nn\\
&+&
e^{-n(\la(x)-\la(x'))} \left(
\frac{\phi_a \cos ( n \phi_a)}{\pi n}
-\left(\frac{\la(x)-\la(x')}{\pi n}+\frac{1}{\pi n^2}\right)\sin( n \phi_a)
\right)
\nn\\
&+&
e^{-n(\tla(x)-\tla(x'))} \left(
\frac{\phi_a \cos ( n \phi_a)}{\pi n}
-\left(\frac{\tla(x)-\tla(x')}{\pi n}+\frac{1}{\pi n^2}\right)\sin( n \phi_a)
\right)
\nn\\
&+&
2e^{-n(\tla(x)-\la(x'))} \left(
\frac{\phi_b \cos ( n \phi_b)}{\pi n}
-\left(\frac{\tla(x)-\la(x')}{\pi n}+\frac{1}{\pi n^2}\right)\sin(n \phi_b)
\right)
\nn\\
&+&
\left.
2e^{-n(\la(x)-\tla(x'))} \left(
\frac{\phi_b \cos ( n \phi_b)}{\pi n}
-\left(\frac{\la(x)-\tla(x')}{\pi n}+\frac{1}{\pi n^2}\right)\sin( n \phi_b)
\right)
\right] .
\label{continuum}
\eea
where we defined a shorthand notation $\phi_a\equiv 2\pi(1-\Da),\phi_b\equiv2\pi(1-\Db)$.

Eq.\eqref{continuum} looks complicated, but
if we integrate by parts repeatedly we can obtain an expansion in terms of $1/N^\alpha$. Consider for instance
\bea
\lefteqn{\int^x_{-x_*} dx' e^{-n(\la(x)-\la(x'))} F(x') }\nn\\
&=&
\int^x_{-x_*} dx' e^{-n N^\alpha (x-x')} e^{-in(y(x)-y(x'))} F(x')
\nn\\
&=&
\frac{1}{nN^\alpha}
\left[ e^{-n N^\alpha (x-x')-in(y(x)-y(x'))} F(x')
\right]_{x'=-x_*}^{x'=x} + {\cal O}(1/N^{2\alpha})
\nn\\
&=&
\frac{1}{nN^\alpha} F(x)  + {\cal O}(1/N^{2\alpha}) .
\eea
The above approximation is valid if in $F(x)$, $x$'s do not appear with a factor of $N^\alpha$.  For instance the density function $\rho(x)$
satisfies this criterion.
In this way we can remove the integration over $x'$ in \eqref{continuum}, and arrive at an expression
\be
F_{int} = N^{2-\alpha} \int^{x_*}_{-x_*} dx \rho(x)^2 f(2y(x)) ,
\ee
where the computation leads to
\bea
\lefteqn{f(2y(x)) = 4 \sum_{n=1}^{\infty}
\Bigg\{ \frac{1}{n^2}
- \frac{\phi_a}{2\pi n^2} \cos(n \phi_a)+\frac{1}{\pi n^3} \sin(n\phi_a)}
\nn\\
&&
 -\frac{\phi_b}{\pi n^2} \cos(n \phi_b) \cos(2ny)
+  \frac{2y}{\pi n^2} \sin(n \phi_b) \sin(2ny)
 +\frac{2}{\pi n^3} \sin(n \phi_b) \cos(2ny)
\Bigg\}
\eea

One can sum this expression using Fourier series of simple linear or quadratic functions. The answer can have a discontinuity at $y=\phi_b/2$ in principle, but rather miraculously, the discontinuities cancel out and we obtain the following simple result.
\be
f(2y) =
\frac{1}{\pi} \phi_b((2\pi-\phi_b)^2-4y^2  ) ,  \quad -\pi < y < \pi .
\ee
Apart from the fact that the intercept at $y=0$ is a function of R-charge assignment, $f(2y)$ being a quadratic function is exactly the same as the result of ABJM model in large-$N$ limit reported in \cite{Herzog:2010hf}.

For the extremization of the whole action we also consider the external force described by the Gaussian part. It is given as
\bea
F_{ext} &=& -i \frac{k}{4\pi} \sum_{i=1}^{N} ( \lambda^2_i - \tilde\lambda^2_i )
\nn\\
&=&  \frac{kN^{1+\alpha}}{\pi} \int^{x_*}_{-x_*} dx x \rho(x) y(x)
\eea
Comparing $F_{ext}$ against $F_{int}$, it is clear the sum should be extremized when $\alpha=1/2$, which is consistent with numerical analysis.

\begin{figure}
\centering
\includegraphics[scale=0.68,trim= 0 0 0 0,clip=true]{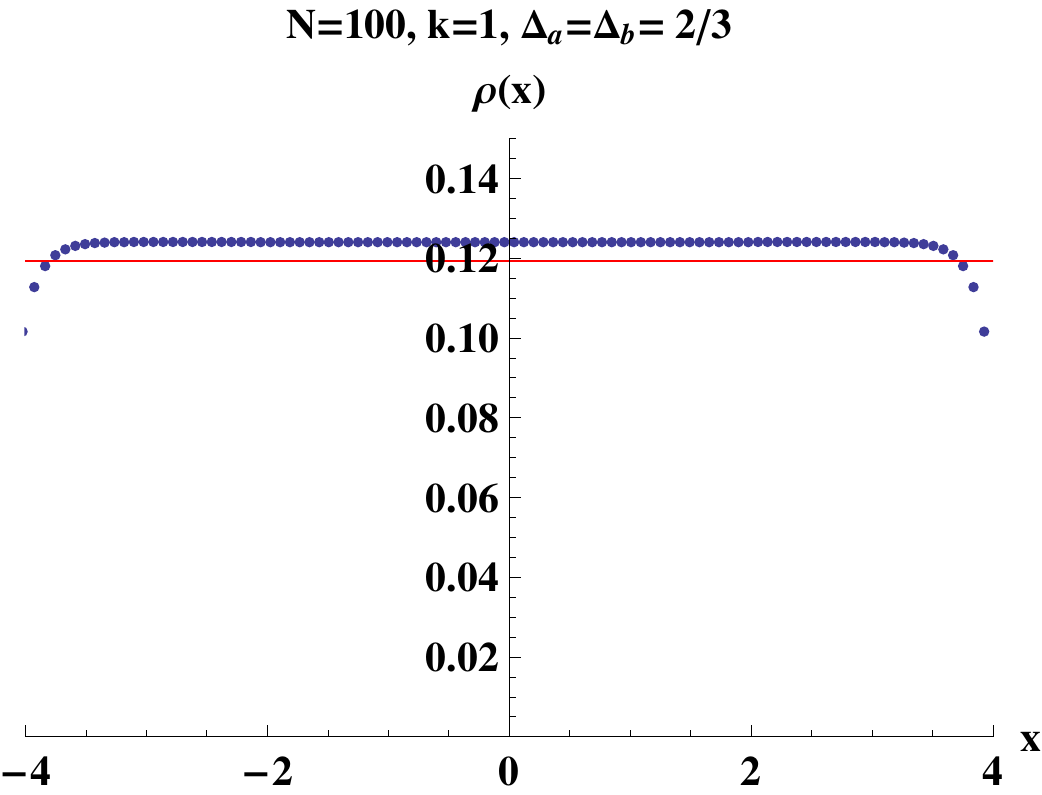}
\includegraphics[scale=0.68,trim= 0 0 0 0,clip=true]{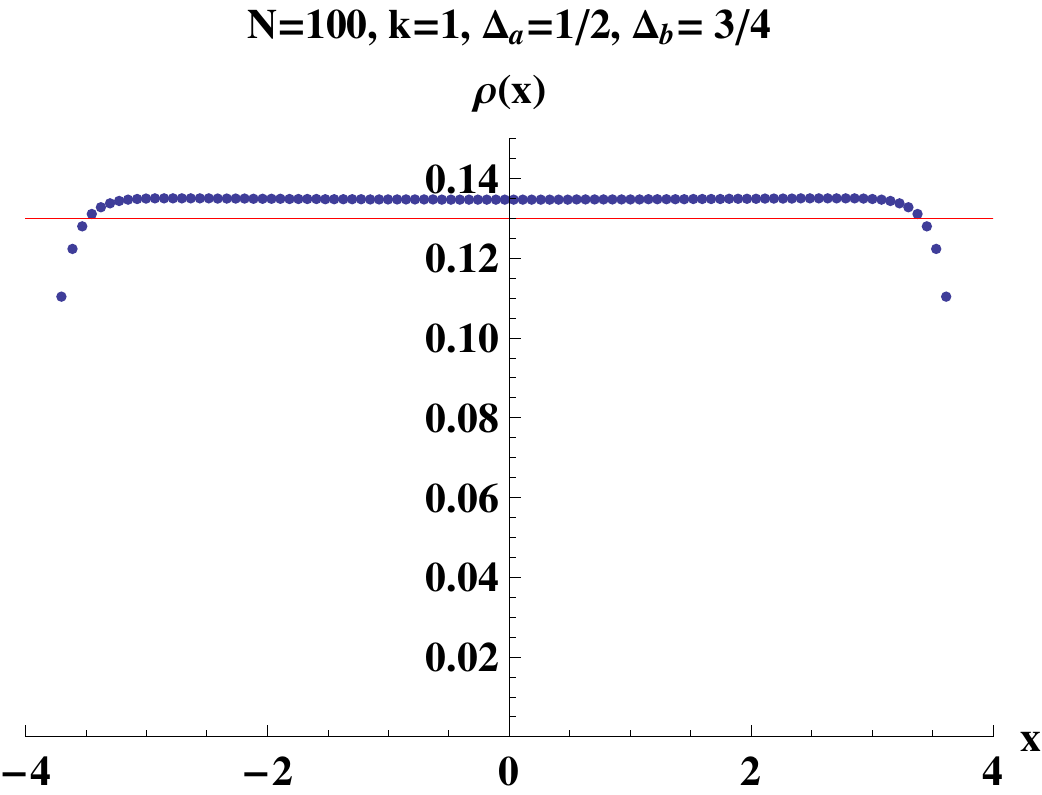}
\caption{These figures show the density of roots $\rho(x)$
for $V^{5,2}$. The dots are numerical data and the red lines show the
analytical predictions.}
\label{rhoV52}
\end{figure}

Taking into account the constraint that the total number of roots is $N$, we have the following expression
\be
F = N^{3/2} \left[
\frac{k}{\pi} \int dx x \rho(x) y(x) + \int dx \rho(x)^2 f(2y(x)) - \frac{\mu}{2\pi}
\left(
\int dx \rho(x) - 1 \right)
\right] \, , 
\ee
where $\mu$ is a Lagrange multiplier.
In fact this is identical to Eq.(16) of \cite{Herzog:2010hf}, and holds in general for any non-chiral models with two nodes we will consider in this paper. It is easy to solve the equation of motion, and one finds the density function is constant while $y(x)$ is linear, i.e.
\be
\rho (x)  = \frac{\mu}{4\phi_b(2\pi-\phi_b)^2} \,  , \quad \quad
y(x) = \frac{k(2\pi-\phi_b)^2}{2\mu} x \, .
\ee
One can check if $\rho(x)$ is really a piece-wise constant function or not, with numerical data. It is indeed the case, as one can see in Fig.\ref{rhoV52}.

Then $F$ is expressed as a function of $x_*$,
\be
F = N^{3/2} \Big[ \frac{\phi_b(2\pi-\phi_b)^2}{2\pi x_*} + \frac{k^2 x_{*}^{3}}{24 \pi \phi_b}  \Big].
\ee
The extremization point is easily found, and with the corresponding imaginary part it is
\be
x_* = \sqrt{ \frac{2\phi_b(2\pi-\phi_b)}{k} },\quad
y_* = \frac{1}{2}(2\pi - \phi_b) = \pi \Db .
\ee
Here we see that $\mbox{Max}(\mbox{Im}(y))$ is indeed bounded by $\pi\Db$, in consistent with our {\it wall of infinite force} argument around Eq.\eqref{force}.

Finally the answer for free energy is
\be
F = k^{1/2} N^{3/2} \frac{\sqrt{2\phi_b (2\pi-\phi_b)^3}}{3\pi} \, .
\ee
Now if we substitute the natural value $\Da=\Db=2/3$, the free energy becomes
\be
F = \frac{16\pi}{27}  k^{1/2} N^{3/2} ,
\label{v52f}
\ee
and it is certainly consistent with the general formula with the volume of $V^{5,2}$ given in  \cite{Martelli:2009ga}
\be
\rm{Vol} (V^{5,2}) = \frac{27\pi^4}{128} .
\ee

\section{Quiver with fundamental matter: $Q^{1,1,1}$ and $V^{5,2}$}
\label{fundam}
\subsection{$Q^{1,1,1}$}
Our second example is a toric homogeneous
Sasaki-Einstein manifold $Q^{1,1,1}$. This space is a $U(1)$ fibration over $\cp^1\times\cp^1\times\cp^1$, so the isometry group is $U(1)_R\times SU(2)^3$. There are several proposals for the field theory duals of $\ads_4\times Q^{1,1,1}$ background. A four-node quiver is proposed for a specific abelian orbifold $Q^{1,1,1}/\mathbb Z_k$  in Ref.\cite{Franco:2009sp}, which also studied a closely related geometry, $Q^{2,2,2}=Q^{1,1,1}/\mathbb Z_2$. These models are chiral, and although it is straightforward to write the integral formula for their partition functions, we found it difficult to perform both numerical and analytic computations.

\begin{figure}
\centering
\includegraphics[scale=0.35,trim= 20 20 40 40,clip=true]{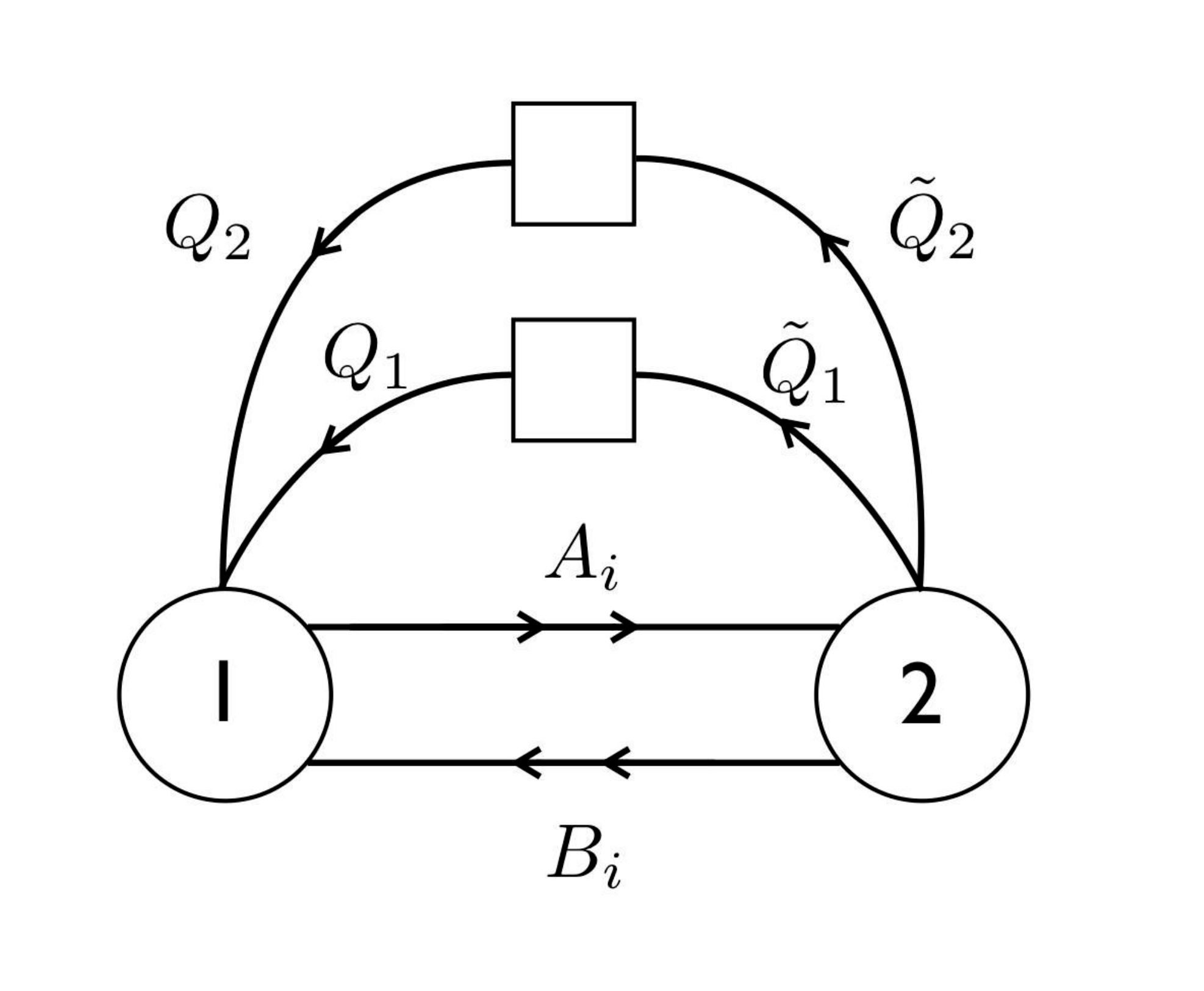}
\caption{The quiver diagram for Chern-Simons dual of $AdS_4 \times Q^{1,1,1}$}
\label{Q111-quiver}
\end{figure}

We consider instead a non-chiral model proposed in \cite{Jafferis:2009th,Benini:2009qs}. Like our previous example, this theory is closely related to ABJM model. In addition to the bifundamental fields $A_i,B_i$, one adds a pair of (anti)-fundamental representation fields to each gauge group node. The quiver diagram is given in Fig.\ref{Q111-quiver}. In addition to the ordinary quartic superpotential of ABJM model, one adds cubic terms
\be
W = \epsilon^{ij}\epsilon^{kl} \,
\Tr A_iB_kA_jB_l + Q_1 A_1 \tilde Q_1 + Q_2 A_2 \tilde Q_2 .
\label{sup2}
\ee

It is argued that this theory gives ${\cal C}(Q^{1,1,1})$ when the bare Chern-Simons levels vanish. Unlike our previous example $V^{5,2}$ in the last section, even if we are using the fact $R(W)=2$ we cannot determine the R-charges completely. Using the $SU(2)\times SU(2)$ symmetry, we may set $R(Q_i)=\Dq, R(\tilde Q_i)=\Dqt$, $D(A_i)=\Delta_A$, and $D(B_i)=\Delta_B$. Using the fact $W$ is a marginal operator, we may expect $\Dq+\Dqt+\Delta_A=2\Delta_A+2\Delta_B=2$. If we assume the bifundamental fields retain the canonical conformal dimension $1/2$, we would get $\Delta_A=\Delta_B=1/2,\Dq=3/4$.

We can write down the partition function as follows.
\bea
Z &=& \frac{1}{(N!)^2}
\int \prod_{i=1}^{N} \frac{d\lambda_i d\tilde\lambda_i}{(2\pi)^2}
\prod_{i<j}\left(2\sinh \frac{\la_i-\la_j}{2}\right)^2 \left(2\sinh \frac{\tla_i-\tla_j}{2}\right)^2
\nn\\
&&
\times \prod_{i,j} \exp{
\left[ 2\ell(1-\Delta_A+i\frac{\la_i-\tla_j}{2\pi}) +
2\ell(1-\Delta_B-i\frac{\la_i-\tla_j}{2\pi}) \right] }
\nn\\
&&
\times \prod_{i} \exp{
\left[ 2\ell(1-\Dq-i\frac{\la_i}{2\pi}) +
2\ell(1-\Dqt+i\frac{\tla_i}{2\pi}) \right] }  .
\label{Q111pf}
\eea
One can easily check this expression is real-valued, and satisfy the following property.
\be
Z(\Delta_A,\Delta_B,\Dq,\Dqt) = Z(\Delta_A,\Delta_B,\Dqt,\Dq) \, .
\ee

Let us now check the convergence of this integral. As we take one of $\la_i$ to infinity while keeping all others finite, we see the leading behavior is
\be
\exp \left[
\left\{
(N-1)(\Delta_A+\Delta_B-1)-2(1-\Dq)
\right\} \la_i
\right] .
\ee
Then clearly for large $N$ the integral is divergent for $\Delta_A+\Delta_B>1$. $\Delta_A+\Delta_B=1$ is the marginal case we are interested in, and then the convergence depends on the value of $\Dq$: We have convergence if $\Dq<1$. And it is the same
 with $\tla$ and $\Dqt$.

\begin{figure}
\centering
\includegraphics[scale=0.70,trim= 0 0 0 0,clip=true]{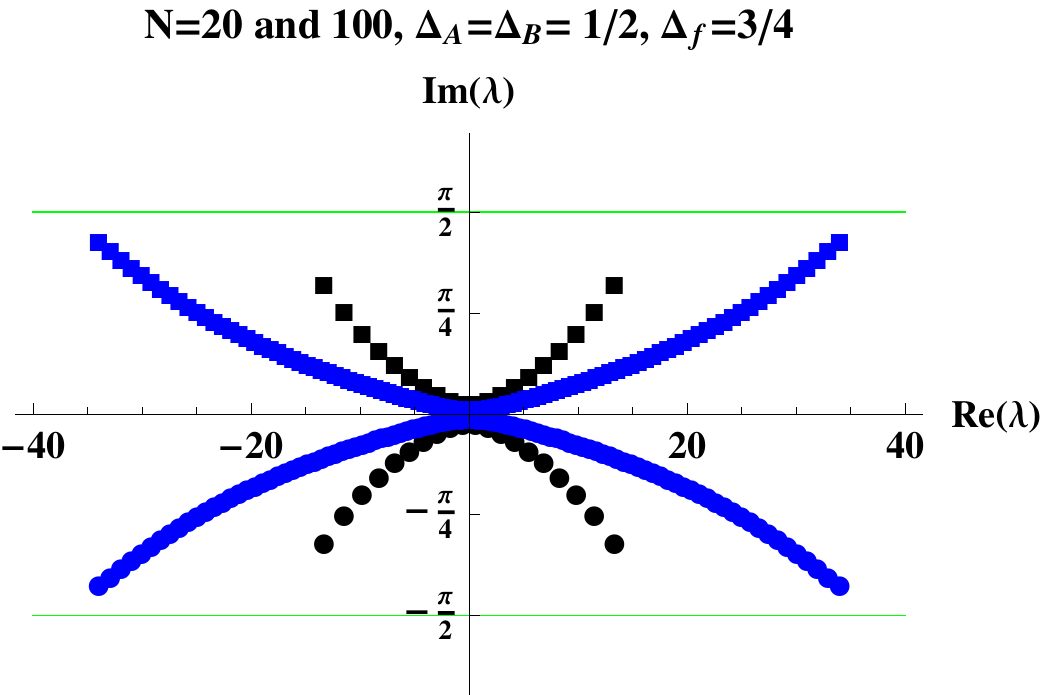}
\includegraphics[scale=0.70,trim= 0 0 0 0,clip=true]{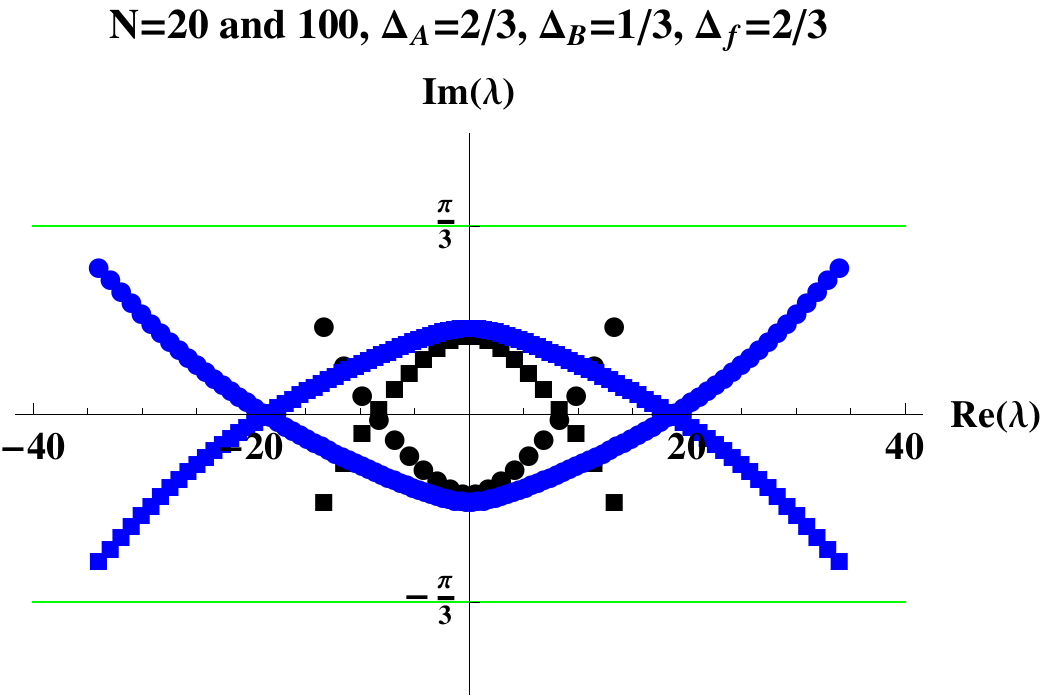}
\caption{The distribution of roots for
$Q^{1,1,1}$ theory with different R charges.
The roots are plotted with black and blue dots for
$N=20$ and $100$, respectively. The real part of the roots
scales as $\sqrt{N}$. The maximum values of imaginary part of
$\lambda$ are bounded by $\pi\Delta_B$, which are represented
by green lines.}
\label{q111roots1}
\end{figure}

With the saddle point equation, one can check that $\la_i^*=\tla_i$ if we assume $\Dq=\Dqt$. In order to simplify the following analysis, we will assume this is the case from now on. The root distribution from the numerical results are given in Fig.\ref{q111roots1}.

\begin{figure}
\centering
\includegraphics[scale=0.63,trim= 0 0 0 0,clip=true]{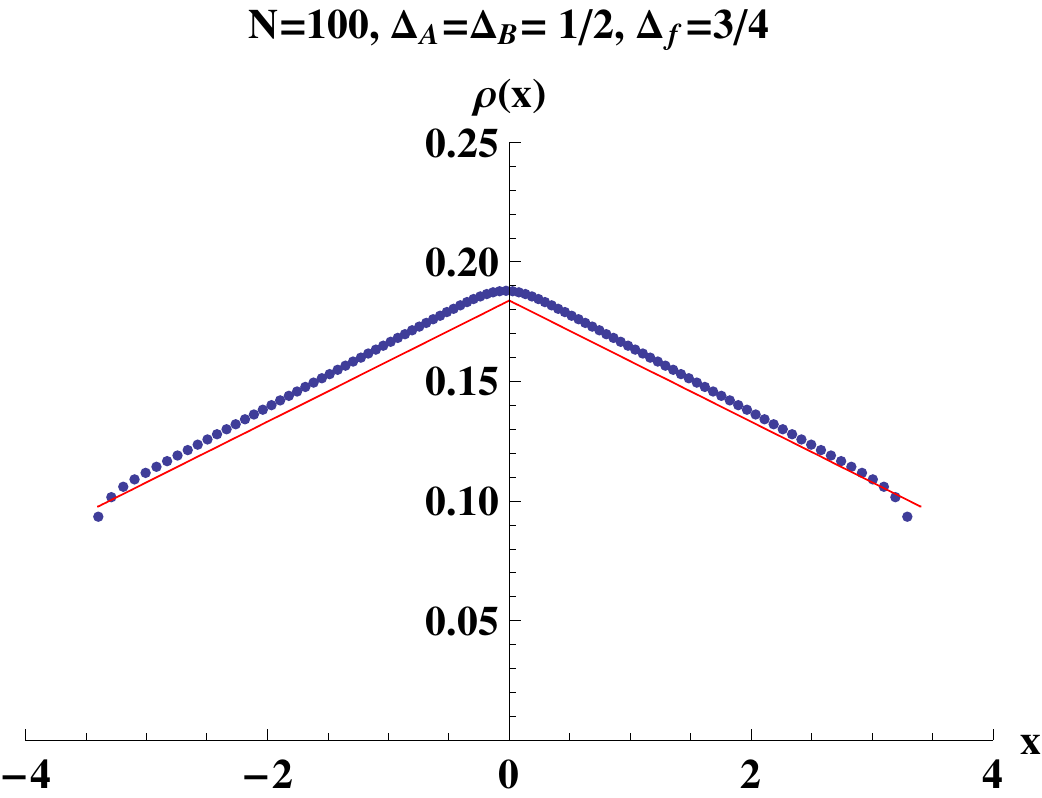}
\includegraphics[scale=0.63,trim= 0 0 0 0,clip=true]{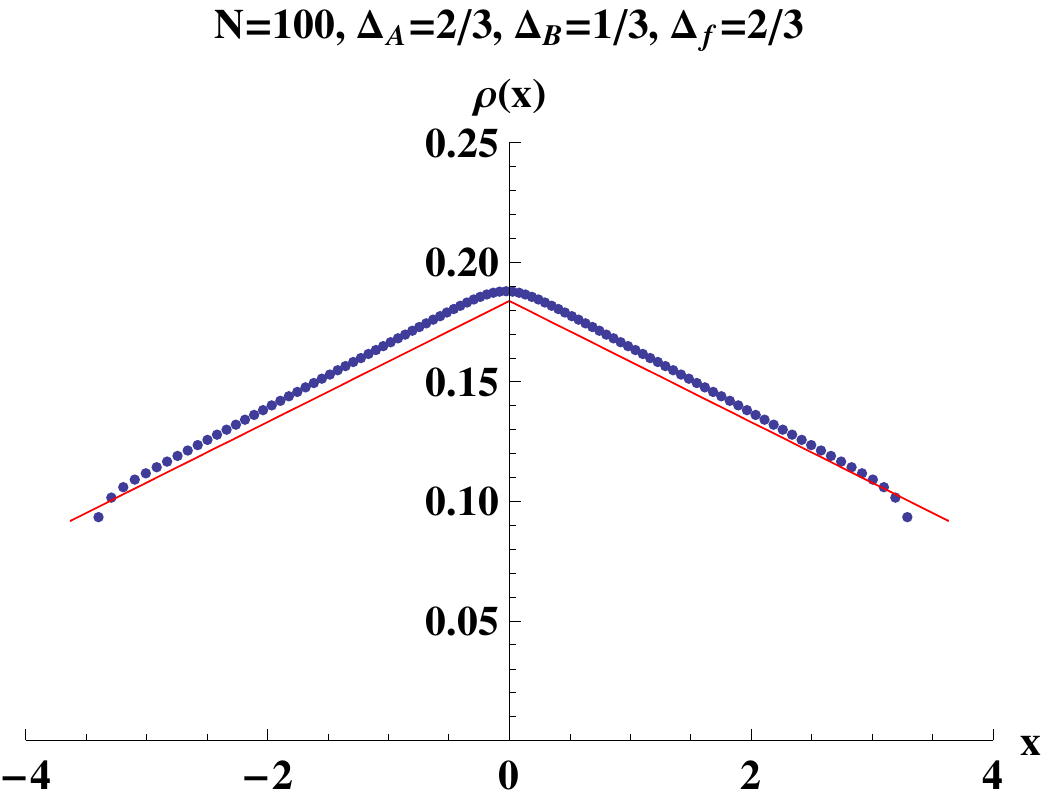}
\caption{These figures show the density function $\rho(x)$
for $Q^{1,1,1}$. The dots are numerical data and the red lines show the
analytical predictions.}
\label{q111d}
\end{figure}




Now we turn to the analytic computation. 
Taking the continuum limit, we first note that the linear terms cancel if $\Delta_A+\Delta_B=1$.
We assume this is true, and after some manipulation similar to Sec.\ref{adj}, we arrive at
\be
F = N^{3/2} \left[\int^{x_*}_{-x_*} dx \rho(x)^2 f(2y(x))
+2  \int^{x_*}_{-x_*} dx \left(1-\Delta_f-\frac{y(x)}{2\pi} \right)\rho(x)|x|  \right].
\ee
The interaction part is described by
\be
f(2y)  = \pi^2 - (2y + 2\pi q)^2
\ee
where we set for convenience $\Delta_A=1/2-q, \Delta_B=1/2+q$. 

Now solving the functional variation equation, we obtain
\be
y=\frac{\pi^2|x|}{(2-2\Delta_f+q)4\pi |x|-2\mu}-q\pi , \quad \quad
\rho= \frac{\mu-(2-2\Delta_f+q)2\pi|x|}{4\pi^3}
\ee
Unlike the previous example, the density function is not constant, and this analytic calculation is supported by numerical results. See Fig.\ref{q111d}.

Then we fix $\mu$ from the normalization condition $\int \rho(x) dx=1$, and find the extremal point of $F$ as a function of $x_*$.
Finally, the free energy as a function of $\Dq,q$ is
\be
F = N^{3/2} \frac{\sqrt{2}\pi(5-4\Delta_f+2q)}{3\sqrt{3-2\Delta_f+q}} \, . 
\ee
If we substitute $\Delta_A+ 2\Delta_f=2$, which is consistent with superpotential
Eq.\eqref{sup2},
\be
F = N^{3/2} \frac{4\pi}{3\sqrt{3}} \, . 
\ee
This matches nicely with the volume of $Q^{1,1,1}$, which can be found for instance in  \cite{Fabbri:1999hw}. 
\be
\mbox{Vol}(Q^{1,1,1}) = \frac{\pi^4}{8} .
\ee
Note that, unlike the previous example of $V_{5,2}$, the right value of $F$ does not uniquely determine the R-charge values for all matter fields.

\subsection{$V^{5,2}$}
We already studied a Chern-Simons-matter theory which describes $V^{5,2}$ in Sec.\ref{adj}.  It is possible to devise another dual field theory using matter fields in fundamental representation \cite{Jafferis:2009th,Benini:2009qs}.  One starts with pure super-Yang-Mills theory without Chern-Simons terms, with gauge group $U(N)$. There are three chiral multiplets $X,Y,Z$ in adjoint representation, and we consider adding $k$ fundamentals $Q_a$ and antifundamentals $\tilde Q_a$, with $a=1,\cdots,k$.

The superpotential is given as
\be
W = \Tr X[Y,Z] + \sum_{a=1}^{k} \tilde Q_a (X^2+ Y^2 + Z^2) Q_a .
\ee
If we are to use this superpotential, the R-charge is obviously $2/3$ for adjoint fields, and $1/3$ for (anti)-fundamental fields. The vacuum moduli space correctly reproduces ${\cal C}(V^{5,2})/{\mathbb Z}_k$  \cite{Jafferis:2009th,Benini:2009qs}.

\begin{figure}
\centering
\includegraphics[scale=0.42,trim= 20 40 40 40,clip=true]{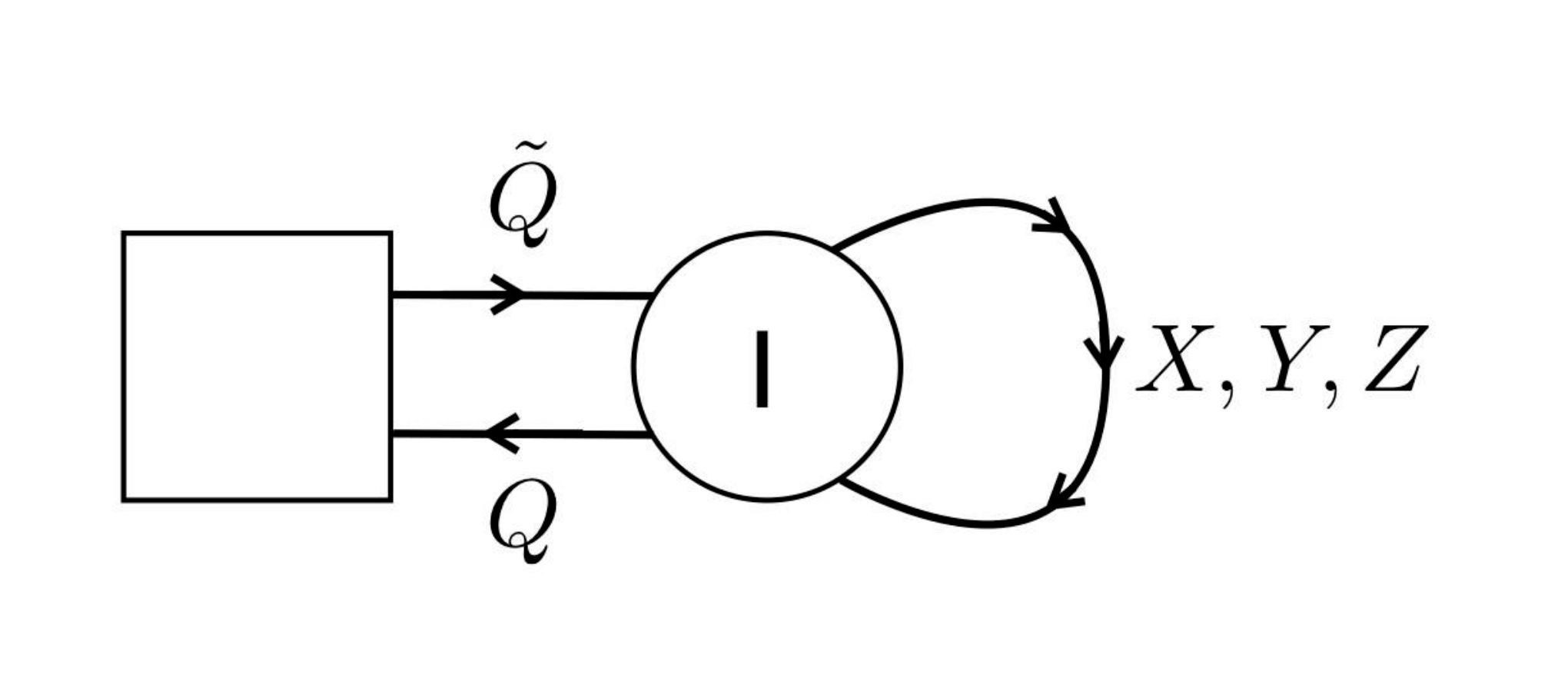}
\caption{The quiver diagram for Chern-Simons dual of $AdS_4 \times V^{5,2}$}
\label{quiver}
\end{figure}


The matrix integral is written as
\bea
Z &=& \frac{1}{N!}
\int \prod_{i=1}^{N} \frac{d\lambda_i}{2\pi}
\prod_{i<j}\left(2\sinh \frac{\la_i-\la_j}{2}\right)^2
\prod_{i,j} \exp{
\left[3 \ell(1-\Da+i\frac{\la_i-\la_j}{2\pi}) \right] }
\nn\\
&&
\times\prod_{i} \exp{
\left[ k\ell(1-\Dq+i\frac{\la_i}{2\pi}) +
k\ell(1-\Dqt-i\frac{\la_i}{2\pi}) \right] }  ,
\label{}
\eea
where $\Da$ is the R-charge of adjoint fields $X,Y,Z$, and $\Dq=R(Q),\Dqt=R(\tilde Q)$.
This expression is always real-valued. 

Let us again consider the convergence of the integral. As we send one particular $\lambda_i$ to infinity while keeping other variables finite, the leading contribution comes from the first line, and the integrand behaves
\be
\exp \left[ \left\{ (3\Da-2)(N-1) - k (2 - \Dq-\Dqt)/2 \right\} \lambda_i  \right]
\ee
So it is obvious that for $\Da>2/3$, the integral in \eqref{mi1} is divergent. As $\Da$ becomes smaller, $Z$ should decay faster and faster. We are basically
interested in the large-$N$ limit with $\Da=2/3$, and perform the computation with an arbitrary R-charge for quarks, i.e. $\Dq=\Dqt\equiv 1-2h$. Note that, once we set $\Da=2/3$, the integral is convergent for $h>0$.

\begin{figure}
\centering
\includegraphics[width=70mm]{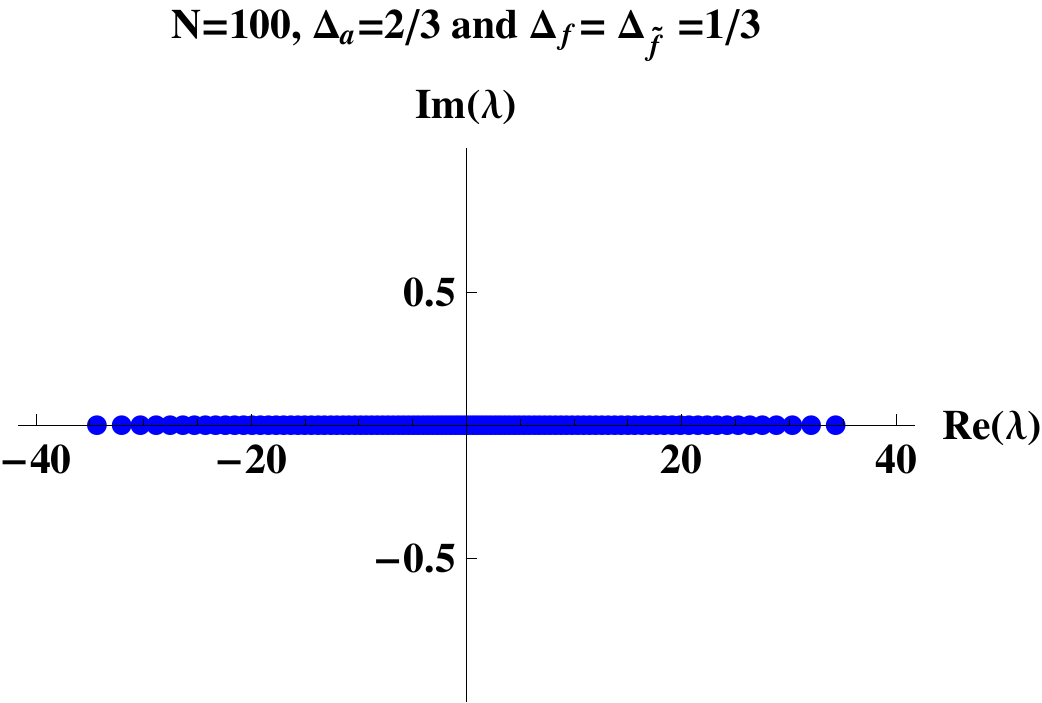}
\includegraphics[width=70mm]{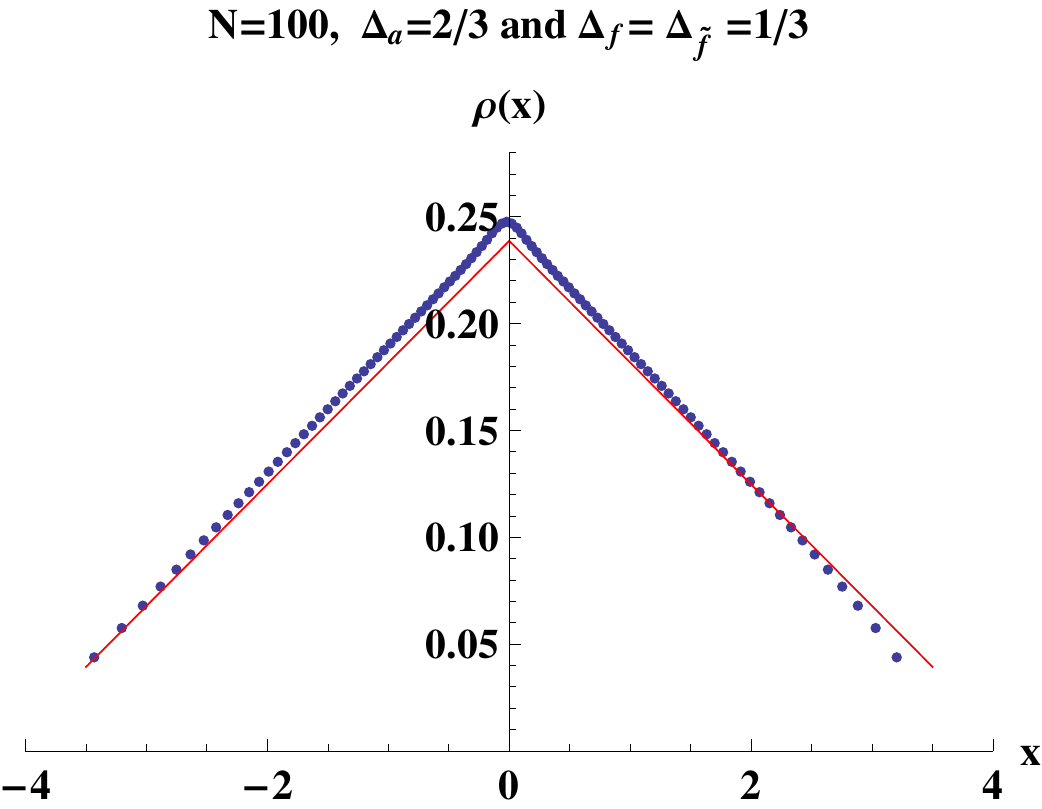}
\caption{The left figure shows saddle points for flavored $V^{5,2}$
theory with $N=100$ and R charge $ \Delta _a=2/3$ and $\Delta _f=
\Delta _{\tilde{f}} =1/3$. All roots are real. The right
figure shows  the density function $\rho(x)$. The dots are numerical data and the red
line shows the analytical predictions. This show that numerical
data and analytical prediction $\rho(x) =3 (4 \pi -3  |x| ) / (16
\pi ^2) $ are same.}
\label{newV52}
\end{figure}

One can proceed in the same way as before, and from the saddle point equation it is clear that the roots are actually real-valued. The linear terms like $\ln e^{\la_i-\la_j}$ get canceled by similar terms from adjoint fields, since $\Da=2/3$.

As the eigenvalues are real-valued at the saddle point, the free energy only contains the density function $\rho$, and the equation of motion in the continuum limit is much simpler than previous examples. From numerical computation we have $\lambda_i = N^\alpha x_i$ and $\alpha \approx 1/2$.  Analytic computation shows $\alpha=1/2$ in the same way as before.

The {\it interaction} part comes from the gauge field contribution and the matter fields in adjoint representation. They are of double-summation form and turn into a multiple integration, but in the large-$N$ limit we can remove one integration by partial integration and evaluating at the boundary.
The final result turns out to be
\be
F_{int} =  \frac{16\pi^2 N^{2-\alpha}}{27} \int^{x_*}_{-x_*} dx \rho(x)^2 .
\ee
Now the {\it quark} fields $Q,\tilde Q$ give single-summation expression in $F$, and in fact they play a similar role as the Chern-Simons terms. In the large-$N$ limit the leading contribution is
\be
F_{fund} = 2hkN^{1+\alpha}  \int^{x_*}_{-x_*} dx \rho(x) |x| .
\ee

Now we set $\alpha=1/2$ and introduce a Lagrange multiplier to write
\be
F = N^{3/2} \left[
 \int dx\frac{16\pi^2}{27}\rho(x)^2 +  2hk\int dx\rho(x) |x|   - \frac{\mu}{2\pi}
\left(
\int dx \rho(x) - 1 \right)
\right]
\ee
From the equation of motion, the density function is obtained
\be
\rho (x)= \frac{27(\mu-4 h k \pi |x| )}{64\pi^3} \, . 
\ee
Note that it is a piece-wise linear function. This fact is confirmed by numerical results, see. Fig.\ref{newV52}.
Now we can determine $\mu$ from the normalization of $\rho(x)$.  The total free energy as a function of $x_*$ is
\be
F = N^{3/2}
\left(
 {hk x_*}-\frac{9 h^2k^2 x_{*}^{3}}{32\pi^2}+\frac{8\pi^2}{27 x_*}
\right)
\ee

The final solution for free energy extremized with respect to $x_*$ is
\be
F = \frac{16\pi}{9\sqrt{3}} (hk)^{1/2} N^{3/2} .
\ee
Now if we use the superpotential and set $\Dq=\Dqt=1/3$ or $h=1/3$, this is the same as \eqref{v52f} and consistent with the volume of
$V^{5,2}$ and AdS/CFT.

\section{Discussion}
\label{discussion}
We have provided a test of $\ads_4/\cft_3$ correspondence for three $\cN=2$ models in this paper. The dual geometries are $Q^{1,1,1}$ and $V^{5,2}$, which are both relatively simple, homogeneous Sasaki-Einstein manifolds.  The partition function is calculated in large-$N$ limit, and in the leading order the free energy with appropriate R-charge assignment shows nice agreement with the predictions of AdS/CFT. Compared with the study involving just the chiral ring structure, our result renders very strong support to the credibility of the dual field theories. 


For explicit computations, we first  used  symmetry argument and marginality of the matrix integral to reduce the number of independent R-charges, and expressed $Z$ as a function of  remaining $R$-charges. 
We can then read off a condition for {\it cancellation of linear terms} from the integrand, like 
$\Delta_A+\Delta_B=1$ in the $Q^{1,1,1}$ model.
 If this condition does not hold, the linear terms with double summation survive and in fact dominate the action. For $\Delta_A+\Delta_B<1$, the integral is convergent and this corresponds to a rather more conventional matrix model integral.  The roots exchange usual $\log |\la_i-\la_j|$ type interaction, and the real parts as well as the imaginary parts are bounded within a finite interval, as $N\rightarrow \infty$. 

\begin{figure}
\begin{center}
\includegraphics[width=75mm]{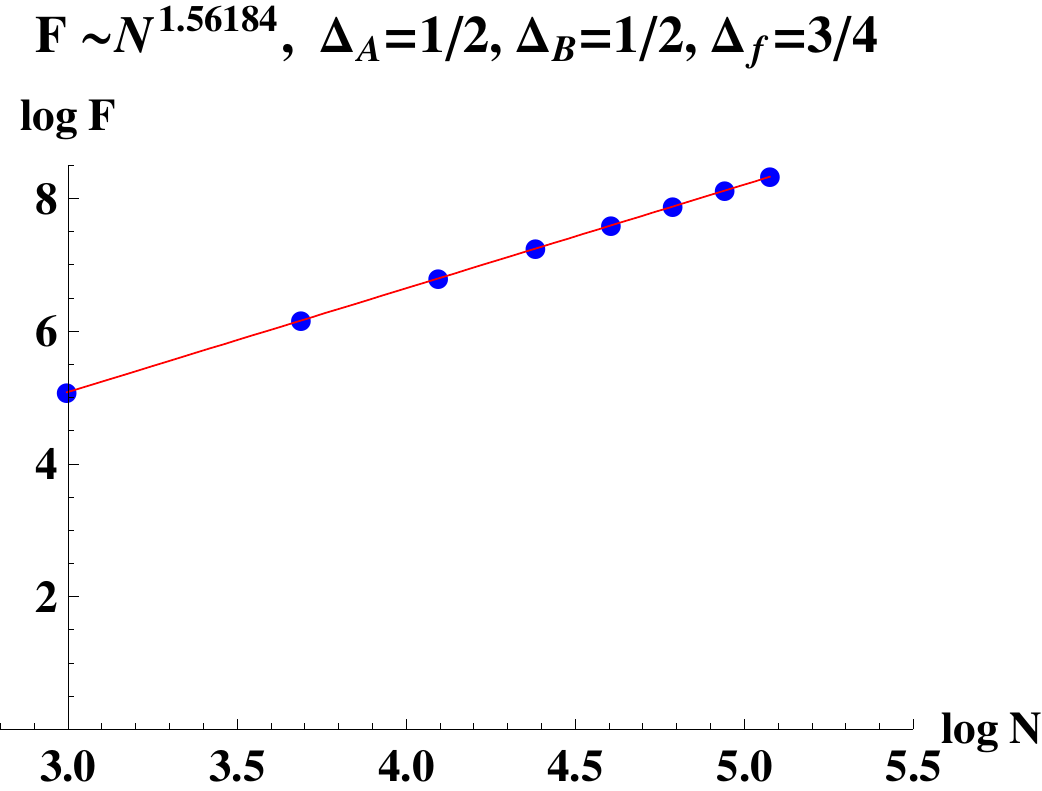}
\includegraphics[width=75mm]{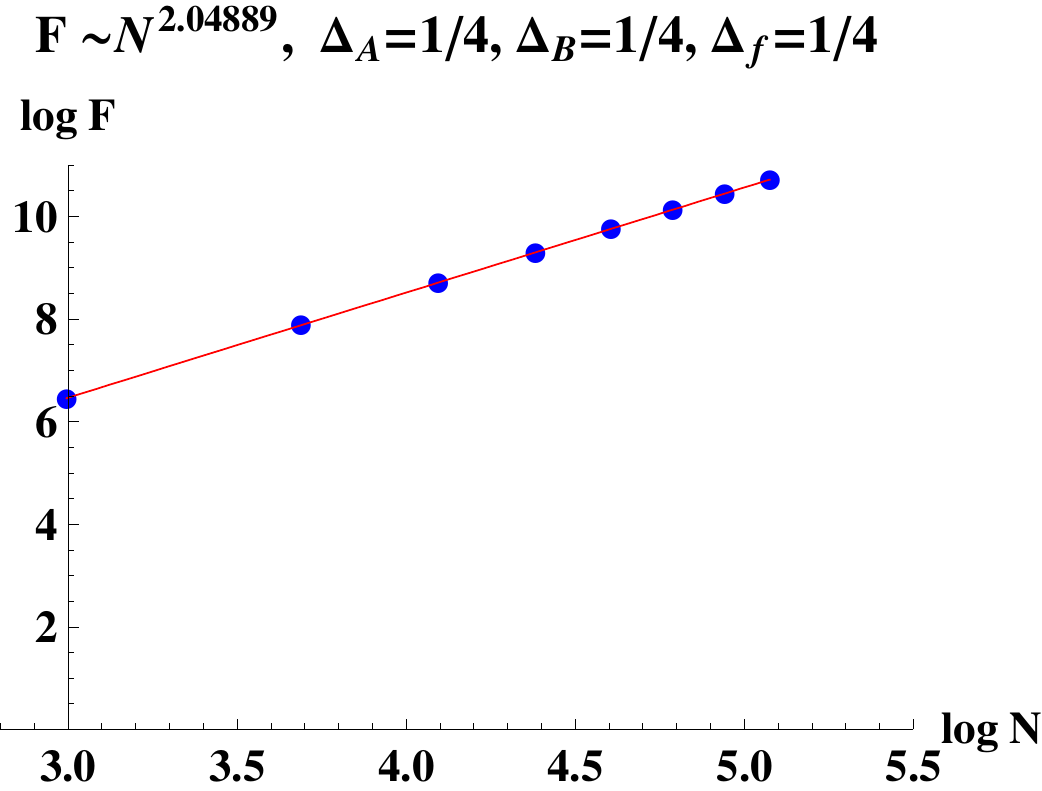}
\end{center}
\caption{These figures how free energy scales with respect to $N$ for different R charges
for $Q^{1,1,1}$ model.
The graphs are log-log plots. The blue dots are numerical data and the red lines are fitting lines.
We took data for  $N=20,40,60,80,100,120,140,160$.
}
\label{Nplots}
\end{figure}

We have checked this prediction numerically. For the $Q^{1,1,1}$ model we have studied here, numerically fitting the data for $N=20,40,\cdots,120$, we have obtained $\alpha=0.569391$ for $\Delta_A=\Delta_B=1/2$, and 
$\alpha=0.104406$ for $\Delta_A=\Delta_B=1/4$.
This makes the free energy scale as $N^2$ below the line of marginality. In Fig.\ref{Nplots} we have illustrated this fact using our numerical data. At a point on the marginal line, $\Delta_A=\Delta_B=1/2$, $F\sim N^{3/2}$ as we have argued already. But away from it, for instance at $\Delta_A=\Delta_B=1/4$, the numerical result gives $F\sim N^2$. If one goes over the line of marginality, with larger R-charges, then $Z\rightarrow\infty$ or $F\rightarrow -\infty$
as we have argued already.

After imposing conditions on R-charges using symmetry and the marginal convergence of $Z$, we  have performed analytic computation and found expressions for $F/N^{3/2}$ as a function of the  remaining R-charges. For the examples considered in this paper, if we 
make full use of the marginality of the superpotential in the dual field theory, $F/N^{3/2}$ agrees exactly with the prediction of AdS/CFT Eq.\eqref{dict}.  

In \cite{Jafferis:2010un}, the author put forward an interesting argument that the partition function as a function of conformal dimensions is extremized at the correct values. In the  large-$N$ limit we are interested in here, $Z$ itself converges to zero and one needs to consider a refinement of $Z$-extremization using ${\cal F} = - \lim_{N\rightarrow\infty} (\ln Z/N^{3/2})$ instead. This is addressed in \cite{Jafferis:2011zi}
\footnote{We thank D. Jafferis who pointed out erroneous statements on $Z$-extremization in earlier versions of this paper.}, where it is shown that the free energy in the large-$N$ limit for strongly coupled ${\cN=2}$ {\it non-chiral} quiver Chern-Simons is consistent with $F$-extremization theorem. 

To check $F$-extremization, one should consider R-charge assignments consistent with the  marginality of superpotential, and then ${\cal F}$ should be extremized by the correct values of the remaining freely-adjustable R-charge values. Our result is in harmony with  $F$-theorem. In fact, when we use  both symmetry argument and marginality of 
superpotential, all R-charge values are either completely determined (for $V^{5,2}$ models) or $F$ has no dependence on the remaining choice of R-charge values (for flavored $Q^{1,1,1}$ model). 

 In order to see the extremization more clearly, for instance with our first example we need to calculate more generally $Z(\Delta_A,\Delta_B,\Delta_{\Phi_1},\Delta_{\Phi_2})$ just using marginality of $W$, i.e. $Z(\Delta_A,\Delta_B=4/3-\Delta_A,\Delta_{\Phi_1}=2/3,\Delta_{\Phi_2}=2/3)$. But it is obvious that this function of $\Delta_A$ should be extremized at $\Delta_A=2/3$, due to symmetry argument.

 In this paper we  studied only non-chiral Chern-Simons quivers with real-valued partition function. For chiral models such as the triangular 
quiver for $M^{3,2}$ model \cite{Martelli:2008si,Hanany:2008cd}, it is more difficult to identify the behavior of roots and perform analytic calculation if we apply the same method used here. We hope to be able to report definite answers for chiral model proposals in future publications.



\acknowledgments
We are grateful to Seok Kim, Sungjay Lee,
Dario Martelli, and Jaemo Park for valuable discussions. We especially
thank D. Jafferis for correspondence and comments on the earlier version of this paper. 
This work was supported by the National Research Foundation
 of Korea (NRF) funded
by the Korean Government (MEST) with the grant number 2009-0085995 (HK,NK),  2010-0023121 (HK,NK),  2010-0007512 (SC),
and also through the Center for Quantum Spactime (CQUeST) of Sogang University
with grant number 2005-0049409 (NK).
\bibliography{pf2}{}
\end{document}